\newif\ifisACM
\isACMtrue

\newif\ifisanon
\isanonfalse

\newif\ifistechreport
\istechreportfalse

\ifisACM
\ifisanon
  \documentclass[sigconf,anonymous]{acmart}
\else
  \documentclass[10pt,sigconf,letterpaper]{acmart}
\fi
\renewcommand\footnotetextcopyrightpermission[1]{} 
\fi

\newcommand{\dns}[1]{{\small \texttt{#1}}}

    \newcommand{\V}[1]{\mbox{\textit{#1}}}  
    \newcommand{\Vs}[1]{\mbox{\textit{\footnotesize #1}}} 
    \newcommand{\Vb}[2]{\V{#1}_{\Vs{#2}}}  

\usepackage{graphicx}
\usepackage{enumerate}
\usepackage{subfig}
\usepackage{amsmath}
\usepackage[linesnumbered,ruled]{algorithm2e}
\usepackage{blkarray}
\usepackage{amsmath}
\usepackage{multicol}
\usepackage{multirow}
\usepackage{hyperref}

\makeatletter
  \newcommand\EatSpacesHack{\@bsphack\@esphack}
\makeatother

\iffalse
  \renewcommand\comment[1]{\textbf{\sffamily\color{blue}[xxx: #1]}}
\else
 \renewcommand\comment[1]{\EatSpacesHack}
\fi

\iffalse
\newcommand\reviewFix[1]{\textbf{\sffamily\color{red}[xxx: reviewFix: #1]}}
\newcommand\PostSubmission[1]{\textbf{\sffamily\color{blue}[xxx: PS: #1]}}
\newcommand\acsacFix[1]{\textbf{\sffamily\color{red}[xxx: acsacFix: #1]}}
\renewcommand\PostSubmission[1]{}
\else
\newcommand\PostSubmission[1]{\EatSpacesHack}
\newcommand\reviewFix[1]{\EatSpacesHack}
\newcommand\acsacFix[1]{\EatSpacesHack}
\fi

\def\Snospace~{\S{}}
\hypersetup{
	colorlinks,%
	citecolor=[rgb]{0.7,0,0},
	filecolor=[rgb]{0.2,0.2,0.2},
	linkcolor=[rgb]{0,0,0.7},
	urlcolor=[rgb]{0.7,0,0},
}

\begin{document}
	
\title{Whac-A-Mole: Six Years of DNS Spoofing}

\ifisACM
\makeatletter
\def\@copyrightspace{\relax}
\makeatother
\ifisanon
  \author{Authors Anonymized for Reivew}
  \acmConference[ACSAC'20 (submission)]{ACSAC'20 (submission)}{Dec. 2020}{Austin, Texas}
\else

\author{Lan Wei \qquad John Heidemann}
\email{{weilan,johnh}@isi.edu}

\affiliation{%
	\institution{University of Southern California/ Information Sciences Institute}
}

\fi
\fi

\begin{abstract}
DNS is important in nearly all interactions on the Internet.
All large DNS operators use IP anycast,
  announcing servers in BGP from multiple physical locations
  to reduce client latency and provide capacity.
However, DNS is easy to \emph{spoof:}
  third parties intercept and respond to queries for benign or malicious purposes.
Spoofing is of particular risk for services using anycast,
  since service is already announced from multiple origins.
In this paper, we describe methods to identify DNS spoofing,
  infer the mechanism being used,
  and identify organizations that spoof from historical data.
Our methods detect overt spoofing and some covertly-delayed answers,
  although a very diligent adversarial spoofer can hide.
We use these methods to study more than six years of data
  about root DNS servers from thousands of vantage points.
We show that spoofing today is rare,
  occurring only in about 1.7\% of observations.
However, the rate of DNS spoofing has more than doubled in less than seven years,
  and it occurs globally.
Finally, we use data from B-Root DNS to validate our methods for spoof detection,
  showing a true positive rate over 0.96.
B-Root confirms that spoofing occurs with both DNS injection and proxies,
  but proxies account for nearly all spoofing we see.
\end{abstract}

\maketitle
\pagestyle{plain}

\section{Introduction}

The Domain Name System (DNS) plays an important part in every web request
  and e-mail message.
DNS responses need to be correct as defined by the operator of the zone.
Incorrect DNS responses from third parties have been used 
  for ISPs to  inject advertising~\cite{Metz09a};
  by governments to control Internet traffic and enforce government policies
  about speech~\cite{Gill15a} or intellectual property~\cite{Crocker11a};
  to launch person-in-the-middle attacks
    by malware~\cite{Cymru14a};
  and by apparent nation-state-level actors
    to hijack content or for espionage~\cite{Krebs19a}.

\emph{DNS spoofing} is when a third-party responds to a DNS query,
  allowing them to see and modify the reply.
DNS spoofing can be accomplished
  by proxying, intercepting and modifying traffic (proxying);
  DNS injection, where responses are returned more quickly than the official servers~\cite{Duan12a};
  or by modifying configurations in end hosts (\autoref{sec:threat_model}).
Regardless of the mechanism,
  spoofing creates privacy and security risks for end-users.

DNSSEC can protect against some aspects of spoofing by insuring the integrity of DNS responses~\cite{Eastlake99b}.
It provides a cryptographic signature that can verify each level
  of the DNS tree from the root.
Unfortunately, DNSSEC deployment is far from complete, 
  with names of many organizations (including Google, Facebook, Amazon, and Wikipedia)
  still unprotected~\cite{Purra14a},
  in part because of challenges integrating DNSSEC with DNS-based CDN redirection.
Even for domains protected by DNSSEC, 
  \emph{client} software used by many end-users fail to check DNSSEC\@.

While there has been some study of how DNS spoofing works~\cite{Duan12a},
  and particularly about the use of spoofing for censorship~\cite{Gill15a},
  to our knowledge, there has been little public analysis of general spoofing of DNS over time.
(Wessels currently has an unpublished study of spoofing~\cite{Wessels19a}.)
Increasing use of DNSSEC~\cite{Eastlake99b},
  and challenges in deployment~\cite{chung2017longitudinal}
  reflect interest in DNS integrity.

\reviewFix{B3:we use percents and so increase in number of probes do not affect things--Lan20190801}
\reviewFix{B5: we should highlight what validation we have, and comment that more is not possible--Lan20190801}
\reviewFix{C1: we should address differentiating mechanism from historical data is our novelty and also providing an overall global view of the DNS spoofing is our novelty. --Lan20190801}
This paper describes
  a long-term study of DNS spoofing in the real-world, filling this gap.
We analyse six years and four months of the 13 DNS root ``letters''
  as observed from RIPE Atlas's
  10k observers around the globe,
  and augment it with one week of server-side data from B-Root to verify our results.
Our first contribution is to define methods to detect spoofing (\autoref{sec:find_spoof})
  and
  characterize spoofing mechanisms from historical data (\autoref{sec:detecting_spoof_mechanisms}).
We define \emph{overt spoofers} and \emph{covert delayers}.
We detect overt spoofers by atypical server IDs;
  they do not hide their behaviors.
We expected to find covert spoofers,
  but instead found 
  covert delayers---third-parties that consistently
  delay DNS traffic but do pass it to the authoritative server.

Our second contribution is to \emph{evaluate spoofing trends} over more than six years of data, 
  showing that spoofing \emph{remains rare} (about 1.7\% observations in recent days),
  but \emph{has been increasing} (\autoref{sec:common}) and is geographically widespread
  (\autoref{sec:where}).
We also identify organizations that spoof (\autoref{sec:who}).

 \acsacFix{A1, A2: methods are heuristic and could have errors. Response: we say although we use heuristic method, our end-to-end check is thorough and we can provide the fact whether the query reaches the B-root.}
Finally, we are the first to validate client-side spoofing analysis
  with server-side data.
We use one week of data from B-Root
  to show that our recall (the true-positive rate)
  is over 0.96.
With the end-to-end check with B-root data, we are able to learn the fact whether or not a query reaches the server.
Server-side analysis confirms that proxying is the most common
  spoofing mechanism.
DNS injection~\cite{Duan12a} and third-party anycast are rare.

\acsacFix{C2: the detection method is straightforward and has been proposed in [16]. Response: I added the following para-Lan20201026}
Our methodology builds on  prior that used hostname.bind queries
  and the penultimate router~\cite{fan2013,jones2016detecting},
  but we provide the first longitudinal study of 6 years of all 13 root letters,
  compared prior work that used a single scan~\cite{fan2013}
   or a day of DNS and traceroute and a week of pings~\cite{jones2016detecting}.
In addition, we are the first to use server-side data to provide end-to-end validation,
  and to classify spoofer identities and eavluate if spoofing is faster.

All data from this paper is publicly available
  as RIPE Atlas data~\cite{Ripe19dns, Ripe19ping, Ripe19trace}
  and from USC~\cite{broot_validate}.
We will provide our tools as open source
  and our analysis available at no cost to researchers.
Since we use only existing, public data about public servers,
  our work poses no user privacy concerns.

\section{Threat Model}
	\label{sec:threat_model}

DNS spoofing occurs when a user makes a DNS query through a recursive resolver
and that query is answered by a third party
  (the \emph{spoofer}) that is not the authoritative server.
We call the potentially altered responses \emph{spoofed}.
We detect \emph{overt spoofers} who are obvious about their identities.
We look for \emph{covert spoofers},
  but find only \emph{covert delayers}
  where DNS takes noticeably longer than other traffic.
We look at reasons and mechanisms for spoofing below.

\subsection{Goals of the Spoofer}
	\label{sec:goals}

A third party might spoof DNS for benign or malicious reasons.

\textbf{Web redirection for captive portals:}
The most common use of DNS spoofing
  is to redirect users to a captive portal so they can authenticate to a public network.
Many institutional wifi basestations intercept all DNS queries
  to channel users to a web-based login page (the portal).
After a user authenticates, 
future DNS traffic typically passes through.

We do not focus on this class of spoofing in this paper because it is transient
  (spoofing goes away after authentication).
Our observers (see \autoref{sec:dataset})
  have static locations (e.g. home) that will not see captive portals.
However, our detection methods would, in principle, detect captive portals
  if run from different vantage points (e.g. hotels).
  
\textbf{Redirecting applications:} 
DNS spoofing can be used to redirect network traffic
  to alternate servers.
If used to redirect web traffic or OS updates,
  such spoofing can be malicious as part of injecting malware
  or exploits.
Alternatively, it can reduce external network traffic.

\textbf{Faster responses:} 
Some ISPs intercept DNS traffic
  to force DNS traffic through their own recursive resolver.
This redirection may have the goal of speeding responses,
  or of reducing external traffic (a special case of redirecting applications,
  or implementing local content filtering (described next).

\textbf{Network Filtering and Censorship:} 
DNS spoofing is a popular method to implement network filtering,
  allowing the ISP to block destinations to enforce local laws
  (or organizational policies, when done inside of an enterprise).
DNS spoofing has been used control pornography~\cite{australia-censorship-bittorrent,australia-censorship-game},
  for political censorship~\cite{Dainotti:2011},
  and to implement other policies.
Spoofing for network filtering
  can be considered a beneficial technique
  or malicious censorship, depending on one's point of view about the policy.
Spoofing for traffic filtering
  can be detected by DNSSEC validation,
  if used.

\textbf{Eavesdropping:}
Since DNS is sent without being encrypted, spoofing can be used to eavesdrop on DNS traffic
  to observe communications metadata~\cite{Farrell14a}.

\subsection{Spoofing Mechanisms}
	\label{sec:spoofmec}

\begin{table*}[t]
\begin{small}
	\begin{center}
		\begin{tabular}{p{2cm}|p{3cm}|p{3cm}|p{3cm}}
			\textbf{mechanism}&\textbf{how} & \textbf{spoofer} &\textbf{spoofee} \\
			\hline
			DNS proxies (\emph{in-path}) & a device intercepts traffic and returns requests & ISPs, universities, corporations & users of the organization \\
			\hline
			\emph{On-path} injection & a device observes traffic and injects responses & hackers, ISPs, governments & anyone whose traffic passes the device \\
			\hline
			Unauthorized anycast site (\emph{off-path}) & a server announcing BGP prefix of the anycast service & ISPs, governments & anyone who accepts the BGP announcement \\
		\end{tabular}
	\end{center}
\end{small}
	\caption{Mechanisms for DNS spoofing.}
	\label{tab:threat}
\end{table*}

\autoref{tab:threat} summarizes three common mechanisms used to spoof DNS: DNS proxies (in-path), on-path injection, unauthorized anycast, following prior definitions~\cite{Duan12a,jones2016detecting}.
We review each mechanism
  and how we identify them in \autoref{sec:detecting_spoof_mechanisms}.

\section{Methodology}
	\label{sec:methodology}
        
\reviewFix{A7: need to tell in a very simple way to making people understand how we differentiate different mechanism --Lan20190731}

We next describe
  our active approach to observe probable DNS spoofing.
This is challenging because, in the worst case, spoofers can 
  arbitrarily intercept and reply to traffic,
  so we use multiple methods (\autoref{sec:find_spoof}).
Moreover, we classify spoofing mechanisms (\autoref{sec:detecting_spoof_mechanisms})
from what we observe from historical data.
Finally, we identify who are the spoofing organizations from the server IDs they returned
  (\autoref{sec:spoofing_party}, and \autoref{tab:classification}).
We caution that our methods are best effort, and not fool-proof against
  a sophisticated adversary.

\subsection{Targets and Queries}
        \label{sec:probing_method}

Our goal is to identify spoofing in a DNS system with IP anycast.
In this paper we study the Root DNS system because it is well documented.
\acsacFix{C4: talking specically about how to generalize}
\comment{this enxt sen is not clear about where you generalize and why ---johnh 2020-11-23}
\comment{I rewrote the following-Lan20201123}
\comment{we should be able to do better than a bullet list ---johnh 2020-11-23}
Our approach can apply to other, non-root DNS anycast systems,
  provided we have access to distributed VPs that can query the system,
  and the system replies to server-id queries (e.g. DNS CHAOS-class), ping, and traceroute.
In principle, our approach could work on anycast systems other than DNS,
  provided they support a query that identifies the server, as well as ping and traceroute.

We probe from controlled \emph{vantage points} (VPs) that can initiate three kinds of queries:
DNS, ping, and traceroute.
We use RIPE Atlas probes for our vantage points
  since they provide a public source of long-term data,
  but the approach can work on other platforms that make regular queries.
In practice, recursive resolvers communicate directly with nameservers on behalf of web clients, 
  so these VPs represent recursive resolvers.

For each VP, we first
  examine basic DNS responses with \emph{Server IDs}
  to detect \emph{overt spoofers} with false-looking server IDs;
Second, we test the timing of replies to
  search for covert spoofers and detect
  \emph{covert delayers}, adversaries who process and forward legitimate replies.
Third, we combine information from all three types of query responses to distinguish the spoofing mechanisms used by the spoofer.

For each hour we observe, we analyze all three datasets (DNS, ping, traceroute).
In that hour, DNS and ping have 15 observations, and traceroute has two observations.
(See how we sample data over time in \autoref{sec:dataset}.).
  
Our targets are authoritative DNS servers using IP anycast.
DNS has three methods to identify server ID:
  CHAOS-class \texttt{hostname.bind}~\cite{Terry:CSD-84-182},
  \texttt{id.server}~\cite{rfc4892},
  and \texttt{NSID}~\cite{rfc5001}.
Each returns a server-specific string, which we call the \emph{Server ID}.
We use \texttt{hostname.bind} because it is supported on all root servers from 2014 to today.

We identify latency via ICMP echo request (ping) to the service address. We also identify penultimate hops of the destination from traceroute.

\subsection{Finding Spoofed DNS responses}
  \label{sec:find_spoof}
  
We examine server ID and ping and DNS latency to identify overt spoofers
  and covert delayers.

\subsubsection{Detecting Overt Spoofers By Server ID}
	\label{sec:overt}
	
We detect covert spoofers because they use Server IDs that differ
  from what we expect.

DNS root operators
  use server IDs that
  follow an operator-specific pattern.
Often they indicate the location, a server number, and the root letter.
For example, A-root operators have a naming convention where the Server ID starts with \textit{nnn1-} and then followed with three letters representing a site/city and end with a number,
  with examples like \emph{nnn1-lax2} and \emph{nnn1-lon3}.
Other root letters follow similar patterns.

By contrast, overt spoofers use other types of names,
  often with their own identities.
Examples include: \emph{sawo}, \emph{hosting}, or \emph{chic-cns13.nlb.mdw1.comcast.net}, \emph{2kom.ru}.

We build a list of regular expressions that match replies from each root operators,
  based on what we observe and known sites as listed at \url{root-servers.org}.
We find server IDs defined by each DNS root operators provide a reliable way to tell spoofing, since our study on years of data shows operators tend to make the server IDs in similar formats across multiple sites.
Also, much fewer vantage points receive atypical server IDs than valid server IDs.
In \autoref{sec:validation}, we prove that recognizing spoofing with atypical server IDs are a reliable way to tell spoofing.

\subsubsection{Detecting Covert Delayers with Latency Difference}
\label{sec:covert}

Although regular expressions can identify spoofers that use
  obviously different Server IDs,
  \emph{covert spoofers} could reused known Server IDs
  to hide their behavior.

We look for covert spoofers
  by comparing DNS and ping latency (as described below),
  assuming that a covert spoofer will intercept DNS but not ICMP\@.
While we find delay differences,
  in all cases, we see that sites with delay difference
  actually pass the query to the authoritative server.
We therefore call what we identify a \emph{covert delayer}.  

Our test for covert delayers
is to compare DNS and ICMP latency
  for sites that have a good-appearing server ID\@.
We compare DNS and ICMP latency by considering all
  measurements of each type for one hour.
\reviewFix{B4: reviewer: some temporary change in the routing a different anycast server could have been chosen. We can just point to your other paper to address that anycast route changes are very rare--Lan20190801 }
(We use multiple measurements to tolerate noise, such as from queueing delay,
  in any given observation.)
For each group of RTTs, we take their median value
  ($\Vb{median}{dns}$, $\Vb{median}{ping}$),
  and median absolute deviations($\Vb{mad}{dns}$, $\Vb{mad}{ping}$) in an hourly window.
We exclude measurements that observe cachement changes
  (based on Server IDs that indicate another location)
  to filter out catchment changes, although we know they are rare~\cite{wei2017does}.
We then define the difference as
  $\delta = |\Vb{median}{dns} - \Vb{median}{ping}|$,
  and require three checks:

\begin{equation}
\left \{
\begin{aligned}
\delta > 0.2 \times \min(\Vb{median}{dns}, \Vb{median}{ping}), \mbox{and}\\
\delta > 3 \times \max(\Vb{mad}{dns}, \Vb{mad}{ping}), \mbox{and}\\
\delta > 10\,\mbox{ms}
\end{aligned} \right.
\label{eqn:criteria}
\end{equation}

The comparison of medians looks for large (20\%), stable differences in latency.
The change also must exceed median absolute deviation
  to avoid overreacting to noisy measurements.
Finally, the check for 10\,ms avoids differences that are around measurement precision.
\acsacFix{[b1d] In Equation 1, it is not clear if these three checks are defined by experience or by experiments. [b1e] Is it possible that attackers can conduct more convert attacks by controlling the latency in order to satisfy these three checks? Response: I added the follow sentences-Lan20201025}
These specific thresholds are based our evaluation of the data,
  and sense that 10\,ms is well beyond normal jitter.
Potentially, a   
  sophisticated adversary could intentionally incrase
  response latency in an effort to bypass these three checks,
  however they cannot reduce latency.

\acsacFix{A8: The paper discusses covert delayers at length, but also mentions that this is not considered as DNS spoofing. It is not clear to me whether those detections are included in the results in Section 4 and 5, then. Response: we say very clear in this section that we do not consider delayers as spoofers-Lan20201025}
\comment{we say they're not spoofers, but YOU DON'T ANSWSER if we omit them from sections 4 and 5.  i propose a sen below, plse look ---johnh 2020-11-11}
While this test is designed to detect covert spoofing,
in practice (details in \autoref{sec:validation})
we see that most cases with large $\delta$ pass the query on to the authoritative server.
The majority of such queries has a larger DNS latency rather than its Ping latency, implying the DNS queries being processed differently by a third-party.
In this paper, we do not consider such interference as DNS spoofing, but consider them as covert delayers.
\comment{clarification added below-Lan20201111}
We therefore count them as valid, non-spoofers in all of \autoref{sec:results}
  and \autoref{tab:common}. 

\subsection{Identifying Spoofing Mechanisms}
\label{sec:detecting_spoof_mechanisms}

Once we detect a spoof,
  next identify the spoofing mechanism
  as anycast or non-anycast (injection or proxy, from \autoref{sec:spoofmec}).

Spoofers can use \emph{anycast} to intercept DNS
by announcing the same prefix as the official DNS servers.
Anycast will affect not only DNS queries,
  but \emph{all} traffic sent to the prefix being hijacked.
Other spoofing mechanisms typically capture only the DNS traffic.

When we look at traceroutes to the site,
  a penultimate hop
  that differs from known legitimate sites
  indicates anycast-based spoofing.
We use the list at
  \url{root-servers.org}
  to identify known sites.
This method is from prior work~\cite{fan2013}.
We consider a VP is under influence of anycast spoof when it meets two conditions.
First, the penultimate hop of its traceroute
should not match that of any VP with an authentic reply,
suggesting the site that the query goes to is not any authentic site.  
Second, its DNS RTT is the same as its Ping RTT,
suggesting in fact anycast is capturing all traffic, not just DNS\@.

\emph{DNS injection} is a second way to spoof DNS~\cite{Dainotti:2011}. 
For DNS injections,
the spoofer listens to DNS queries (without diverting traffic),
then replies quickly, providing an answer to the client before
the official answer sent from an authoritative server.
The querying DNS resolver accepts the first, spoofed reply
  and ignores the additional, real reply.

DNS injection has two distinguishing features:
responses are fast and doubled~\cite{schomp2014assessing}.
Without a platform that preserve multiple responses for one query, it is hard from historical data to recognize injection mechanism.

\emph{Proxies} are the final spoofing mechanism we consider.
A DNS proxy intercepts DNS traffic,
then diverts it to a spoofing servers.
Unlike DNS injection, the original query never reaches the official server because proxy simply drops the queries after returning the answer.

Using historical data collected from VP's side, we can only differentiate mechanism between anycast vs. non-anycast.
With further validation from Root DNS server-side data in \autoref{sec:validation}, we can differentiate mechanisms between injection vs. proxy.

\subsection{Spoofing Parties from Server IDs}
	\label{sec:spoofing_party}

Spoofing is carried out by multiple organizations;
  we would like to know who they are and  identify \emph{unique spoofing parties}.
We use patterns of Server ID to identify overt spoofers,
  and after knowing who they are,  we classify them to seven categories based on their functions.

We identify unique spoofing parties through several steps.
First, for Server IDs that have a recognizable DNS name,
  we group them by common prefix or suffix (for example, \dns{rdit.ch} for
  \dns{njamerson.rdit.ch} and \dns{ninishowen.rdit.ch}).
We handle Server IDs and IP addresses the same way,
  after looking up their reverse DNS name.
We manually identify and group recognizable company names.
Remaining Server IDs are usually generic (DNS13, DNS-expire, etc.), for which we group by the AS of the observing VPs.

We classify identifiable spoofing parties
by examining their websites.
Each class is based on the goal or function of the organization or the person.
\autoref{tab:classification} shows seven class of different spoofing parties (ISPs, DNS tools, VPNs, etc.), with specific examples provided.
Our work maintains a full table of spoofers seen in years and their responding webpages.
An example table is at \autoref{sec:who}, with clickable example URLs showing the identity of spoofing organizations.

\section{Results}
	\label{sec:results}

We next study six years and four months of Root DNS to look for spoofing.
First, we study the quantity of DNS spoofing. 
We show it is uncommon but is getting more popular over time. 
Second we study the locations and identities of the spoofers.
Finally we discuss whether spoofing always provides a faster response than authorized servers.

\subsection{The Root DNS system and Datasets}
	\label{sec:dataset}

We observe the Root DNS system using RIPE Atlas~\cite{ripe}.

\textbf{Background about Root DNS system:} Root DNS is provided by 13 independently operated services, named A-root to M-root~\cite{RootServers19a}.
All of the root letters use IP anycast\footnote{
        H-Root's sites are primary/secondary, so only one is visible on
          the general Internet at a time.
        },
  where locations, typically in different cities, share a single IP address.
The number of locations for each letter varies, from a few
  (2 for H, 3 for B, less than 10 for C, G, and 28 for A)
to hundreds (D, F, J, and L all operate over 100) as in August 2019~\cite{RootServers19a}.
We use the list of anycast locations at \url{root-servers.org}
  as ground truth.

\textbf{RIPE Atlas:}
Our observations use public data collected by RIPE Atlas probes from 2014-02 to 2020-05 (six years and four months).
RIPE Atlas has standard measurements of
  DNS server ID (hostname.bind), ICMP, and traceroute (UDP) to each Root Letter
  for most of this period.
Exceptions are that  
  G-Root never responds to ICMP,
  and E-Root data is not available from 2014-02 to 2015-01.

We show the frequency of each type of query in \autoref{tab:query}.
In each part, we sample at a random one-hour window each of the three type of datasets.
Over the multi-year period, we extract 4 observations each month.
Each measurement is a randomly chosen hour in a different week of the month,
  with the first in the 1st to 7th of the month,
  the second in the 8th to the 14th, then 15th to the 21st,
  and finally 22nd to the end of the month
  (``weeks'' are approximate, with the fourth week sometimes longer than 7 days).
We choose the same hour for all letters,
  but the hour varies in each week to avoid bias due to time-of-day.

\begin{table}
	\begin{center}
		\begin{tabular}{p{1.7cm}|p{1.7cm}}
			\textbf{type}&\textbf{frequency} \\
			\hline
			DNS & every 240s \\
			
			Ping & every 240s  \\
			
			Traceroute & every 1800s\\
		\end{tabular}
	\end{center}
	\caption{Query detail}
	\label{tab:query}
\end{table}

The exact number of VPs we use varies,
  since VPs sometimes disconnect from RIPE Atlas infrastructure (VPs are individually owned by volunteers),
  and RIPE adds VPs over the measurement period.
The number of VPs,  ASes, and countries measured over time is show
  in \autoref{tab:coverage}.
\acsacFix{C3: ripe atlas biased. Response: added the following line-Lan20201026}
Our result is limited by the coverage of RIPE atlas VPs.
  
\begin{table*}
	\begin{center}
		\begin{tabular}{c|rrrrrrr}
			\textbf{Year}& \textbf{2014} & \textbf{2015} & \textbf{2016} & \textbf{2017} & \textbf{2018} & \textbf{2019} & \textbf{2020}\\
			\hline
			\textbf{Vantage Points} &7473 &9223&9431&10311&10336 &10492 & 10988\\
			
			\textbf{AS} &2616&3322 &3370&3633&3605&3590 &3397 \\
			
			\textbf{Country} &168 &184&186&183&181 &180 &175\\
		\end{tabular}
	\end{center}
	\caption{Data Coverage}
	\label{tab:coverage}
\end{table*}

\subsection{Spoofing Is Not Common, But It Is Growing}
	\label{sec:common}
In this section, we talk about how much spoofing occurs today, and what is the trend of spoofing over the six years and four months.

\subsubsection{Spoofing is uncommon}

\begin{table}
	\centering
	\begin{tabular}{l@{}rr}
		&\multicolumn{2}{c}{\textbf{2020-05-03}}\\
		active VPs & 10882 & 100.00\%\\
		\hspace{10pt}timeout & 260 & 2.39\% \\
		\hspace{10pt}answered & 10622 & 97.61\% \\
		\hspace{20pt}valid & 10430 & 95.85\%\\
		\hspace{30pt}covertly-delayed & 19 & 0.17\%\\
		\hspace{20pt}spoofed & 192 & 1.76\%\\
		&&\\
	\end{tabular}
	\caption{DNS spoof observations.}
	\label{tab:common}
\end{table}

Spoofing today is uncommon: about 1.76\%, 192 of the of 10882 responding VPs are spoofed.
\autoref{tab:common} shows on 2020-05-03: that about 95.85\% of all VPs received valid answer (in which totally 19 (0.17\%) VPs experience delayed DNS answers), 1.76\% VPs are overtly spoofed.
More VPs (2.39\%) timeout than see spoofing.

\begin{figure}
 	\begin{center}
 		\includegraphics[width=0.8\columnwidth]{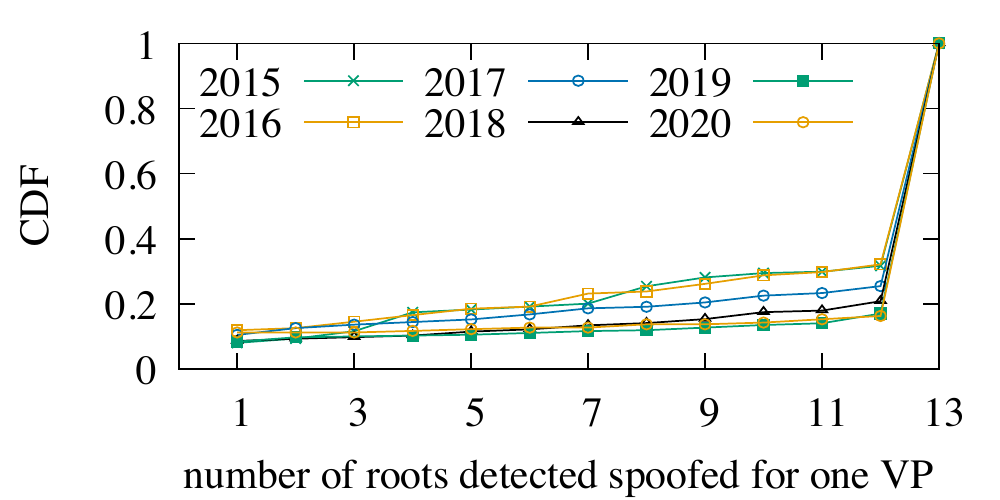}
 	\end{center}
 	\caption{CDF of root counts seen overtly-spoofed (2014 not provided because of lacking E-root data)}
 	\label{fig:rootcount}
 \end{figure}

Most spoofers spoof on \emph{all} root letters.
Fig.~\ref{fig:rootcount} shows the CDF of how many letters are spoofed
  for each VP, and
  of VPs that see spoofing, there are always more than 70\% of VPs see spoofing of all root letters throughout the six years we observed.
  In year 2020, there are 83\% of VPs experience spoofing across all root letters.
 
\subsubsection{Growth}
\begin{figure*}
	\begin{minipage}[b]{0.8\textwidth}
		\includegraphics[width=\columnwidth]{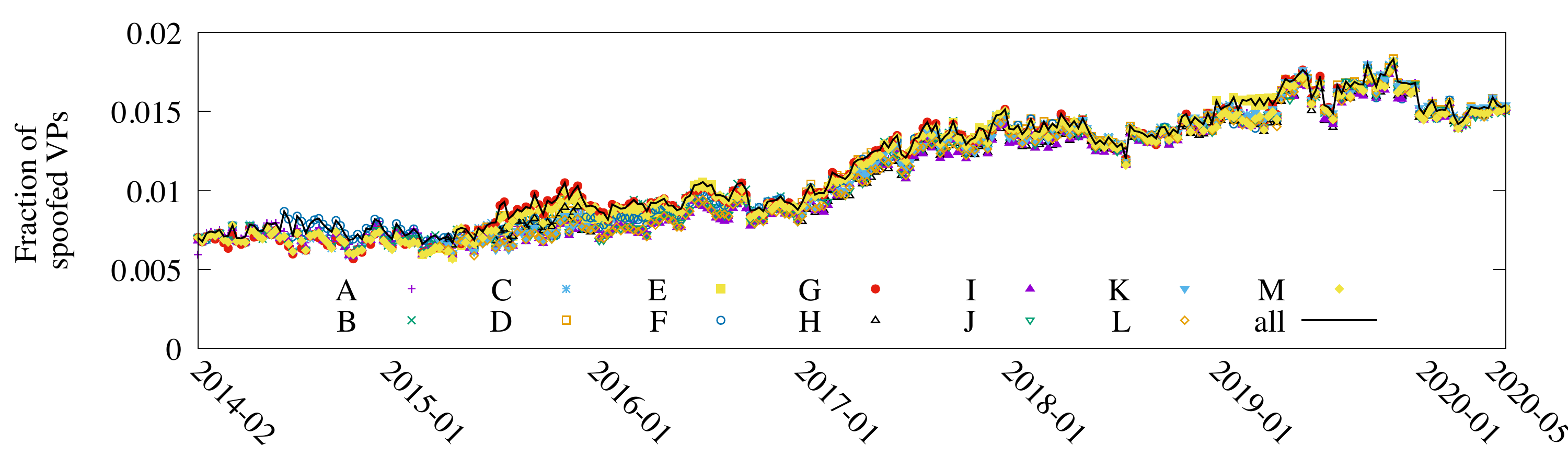}
		\caption{Fraction of spoofed VPs over all available ones at each date.}
		\label{fig:trend_all}
	\end{minipage}
	\begin{minipage}[b]{0.19\textwidth}
		\includegraphics[width=\columnwidth]{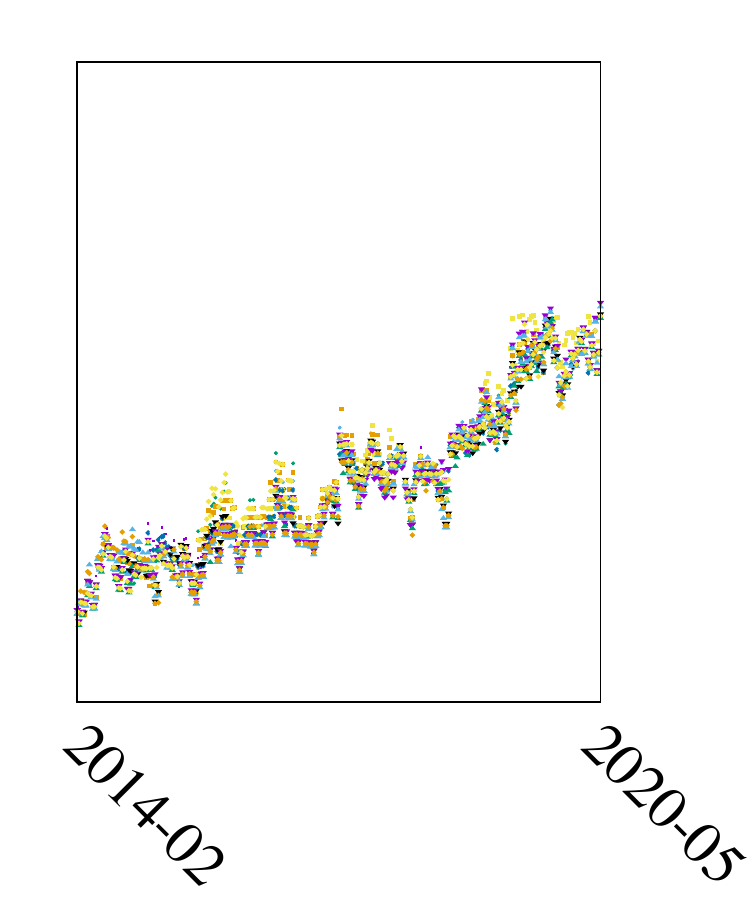}
		\caption{3000 VPs}
		\label{fig:trend_fix_group}
	\end{minipage}
\end{figure*}

We see an increasing amount of spoofing over the six years and four months we study.
\autoref{fig:trend_all} shows the fraction of VPs that see \emph{any} root servers spoofed (the thick black line), and for each root letter spoofed (colorful dots).
Although we see some variations from day to day, the overall fraction of spoofed VPs rises from 0.007 (2014-02-04) to 0.017 (2020-05-03), more than doubling over six years.

Because the set of active VPs changes and grows over time,
  we confirm this result with a fixed group of 3000 VPs
  that occur most frequently over the six years,
  shown in \autoref{fig:trend_fix_group}.
This subset 
  also increased to more than twice over the six years, but from a slightly lower baseline  (0.005) to 0.014,
  confirming our findings.
In \autoref{sec:where}, we later show that location affects the absolute fraction of spoofing.

\subsection{Where and When Are These Spoofers? }
\label{sec:where}

\begin{figure*}
	\centering
	\centering
    \includegraphics[width=0.95\columnwidth]{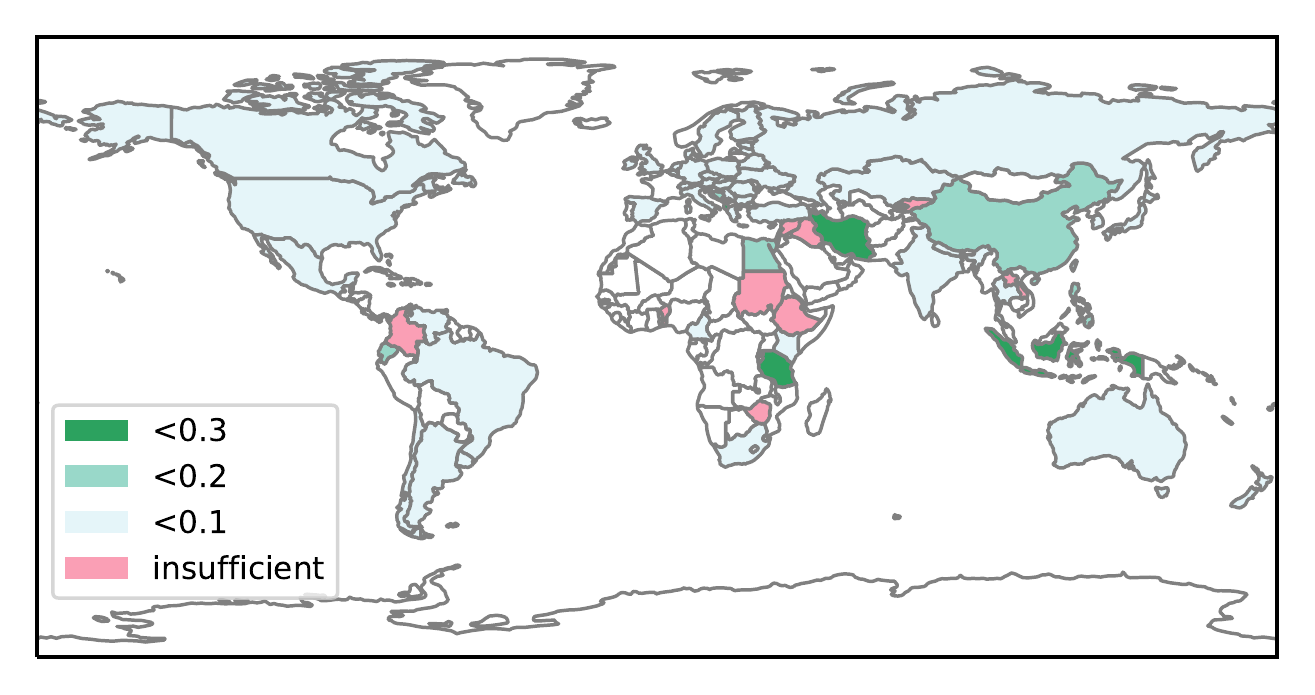}
    \caption{Fraction of spoofing per country (varied green shades), spoofed with under-sampled VPs (pink), not spoofed (white).}
    \label{fig:density_map}
\end{figure*}

\begin{figure*}
	\begin{center}
	\includegraphics[width=0.95\columnwidth]{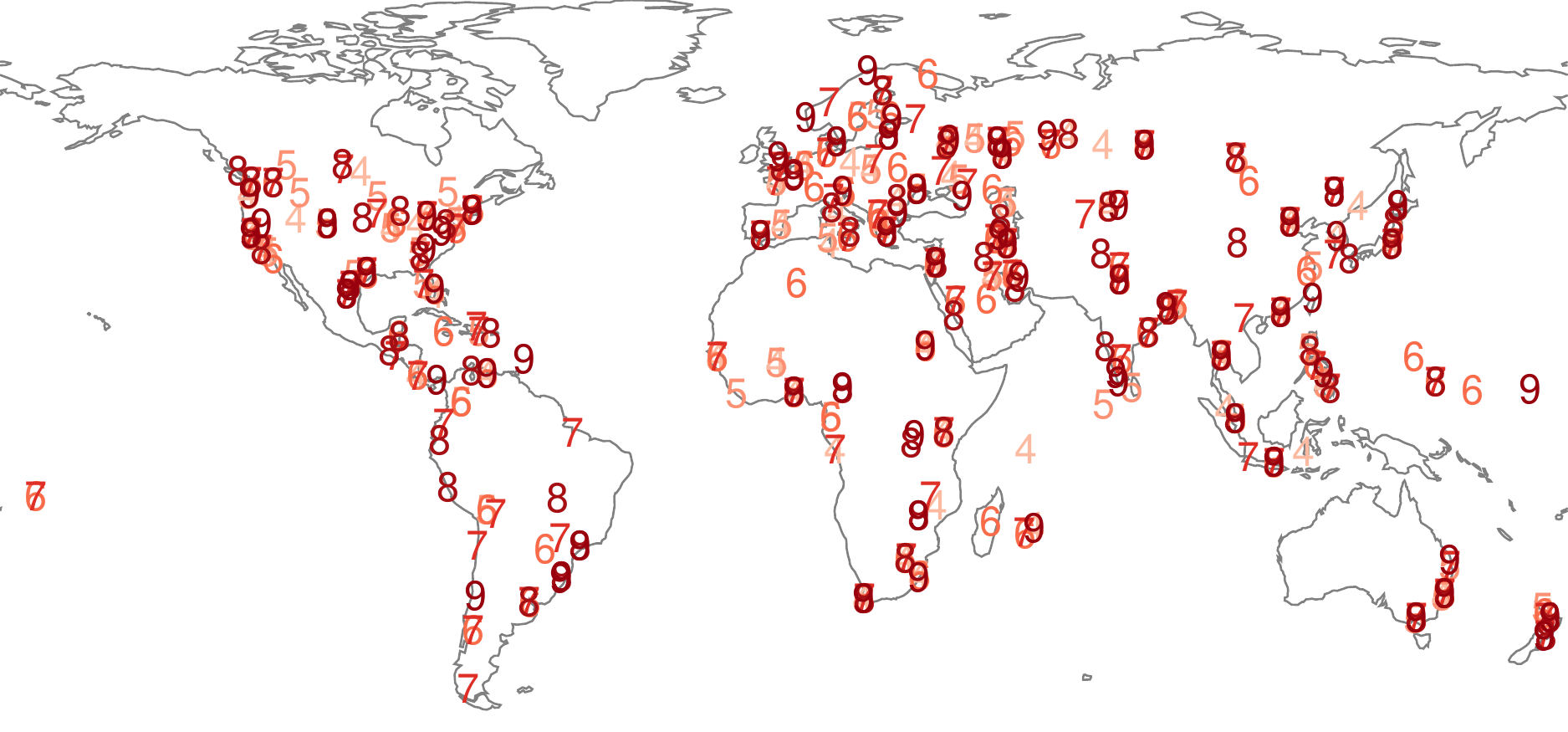}
    \end{center}
     \caption{Location of spoofed VPs, slightly jittered.  More recent years are darker colors, and each loation is the last digit of the year of the observation. Overlapping digits indicate spoofs over multiple years.}
\label{fig:timelocation}	
\end{figure*}

\reviewFix{One reviewer asks for a histogram over countries instead of maps, i think maybe other reviewers prefer maps, since maps show that spoofing really happens in all continents. But for the top 10 countries, do we want to provide a table for the data, I put them in text in the following para--Lan20191022}

We next consider \emph{where} spoofing happens.
If spoofing is legally required in some countries,
  we expect spoofing to be concentrated there.

\autoref{fig:density_map} shows the fraction of VPs that see spoofing, by country
  (countries with less than 10 active VPs but are spoofed are listed as ``insufficient'' and are excluded from our ranking).
From 2019-01 to 2019-08, we see spoofing is most common in the Middle East and Eastern Europe, Africa, and Southeast Asia.
  but we see examples of spoofing worldwide.
The top ten countries by fraction of spoofing is in \autoref{tab:topten}.
\begin{table}
	\centering
	\begin{tabular}{lrrr}
		& \multicolumn{2}{c}{\textbf{VPs}} & \\
		\textbf{Country} & \textbf{spf.} &\textbf{active} & \textbf{\%} \\    
		Indonesia& 23&87& 26\\ 
		Iran&48&198& 24\\ 
		Tanzania& 2& 10& 20\\ 
		Albania& 8& 40& 20\\ 
		Philippines& 4& 26& 15\\ 
		Ecuador& 2& 15& 13\\ 
		Bosnia \& Herz& 2& 18& 11\\ 
		China& 3& 27& 11\\ 
		Egypt& 1& 10& 10\\ 
		Lebanon& 2& 20& 10\\ 
		&&&\\
	\end{tabular}
	\caption{Countries with largest fraction of VPs experiencing spoofing in 2019.}
	\label{tab:topten}
\end{table}

Most areas show spoofing activity over multiple years.
\autoref{fig:timelocation} shows our six years (without year 2020) of spoofing occurrence
  with different years in its last digit as symbols and in different darknesses.
(Points in oceans are actually on islands.).
\acsacFix{A7: labels overlap with each. Response: we say what we overlap them for.-Lan20201025}
Labels that overlap show VPs that are spoofed multiple times over different years.

\subsection{Who Are the Spoofing Parties?}
\label{sec:who}

\reviewFix{A5: reviewer would like to see reflections between goals in section 2.1 and our results. Here would be a good place to add reflections.--Lan20190731 }

Goals of spoofers (\autoref{sec:goals})
  include faster response, reduced traffic, or censorship.
With more than 1000 root instances,
  a strong \emph{need} for spoofing for performance seems unlikely,
  although an ISP might spoof DNS\@.
We next study the identification  of spoofing parties and classify them to perhaps infer their motivation by using the methodology in \autoref{sec:spoofing_party}.

\begin{table}
	\centering
	\begin{tabular}{p{1.9cm}|p{2.6cm}|p{1.9cm}}
		\textbf{Types} & \textbf{Example URLs} &\textbf{Number of \newline clustered spoofers}\\
		\hline
		ISPs & \url{skbroadband.com}\newline \url{2kom.ru} &32~(16.16\%)\\
		\hline
		network \newline providers& \url{softlayer.com}\newline \url{level3.com}&24~(12.12\%)\\
		\hline
		education-\newline purpose & \url{eenet.ee} &1~(0.5\%)\\
		\hline
		DNS tools & \url{dnscrypt.eu} &1~(0.5\%)\\
		\hline
		VPNs & \url{nordvpn.com} & 1~(0.5\%)\\ 
		\hline
		hardware & \url{eero.com} & 1~(0.5\%)\\
		\hline
		personal & \url{yochiwo.org} &1~(0.5\%)\\ 
		\hline
		unidentifiable & DNS13\newline DNS-Expire&137~(69.19\%)\\
	\end{tabular}
	\caption{Classification of spoofing parties.}
	\label{tab:classification}
\end{table}

\autoref{tab:classification} shows the spoofing parties we found.
More than two-thirds, spanning 137 ASes,
  show generic Server IDs and are unidentifiable.
Of identifiable spoofers, most are end-user (eyeball) ISPs (32, about half)
  or network providers (24 providing cloud, datacenter, or DNS service).

Sometimes spoofing parties do not affect all VPs in the same AS.
T-Mobile spoofs a VP in Hungary, but not elsewhere.
In Comcast, 2 of the 322 VPs see spoofing,
  and in FrontierNet, 1 of the 28 VPs sees spoofing.

Identifying spoofing parties suggests possible reasons for spoofing:
  the ISPs may be improving performance,
  or they may be required to filter DNS.
The 5 classes each with 1 example are likely spoofing for professional interest,
  because they work with DNS or provide VPNs.

\subsection{How Do Spoofing Parties Spoof?}

\begin{figure}
		\begin{center}
		\includegraphics[width=0.85\columnwidth]{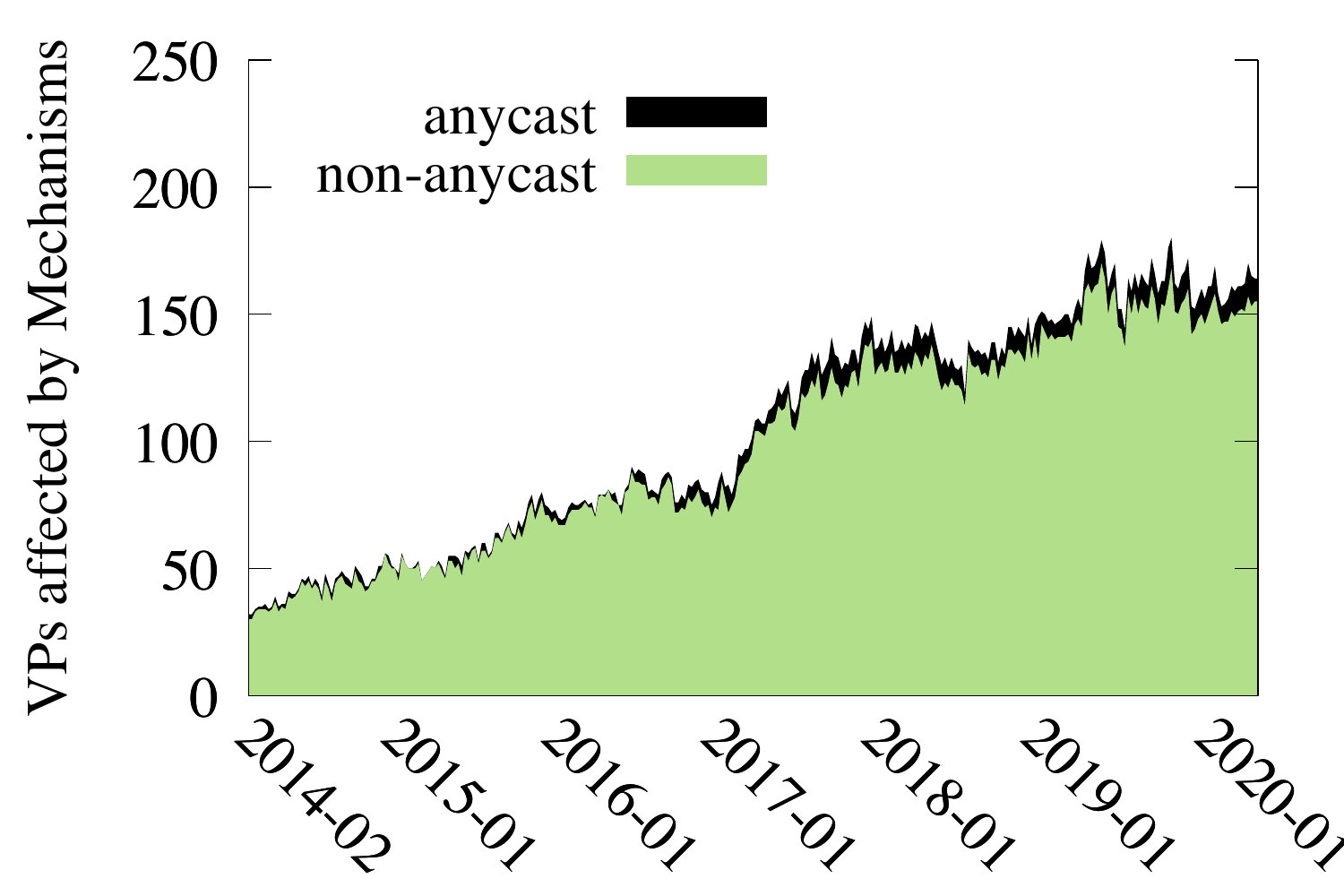}
		\end{center}
		\caption{Number of VPs with different spoofing mechanisms over time.}
		\label{fig:mechanism}
\end{figure}

We next examine spoofing mechanisms, following \autoref{sec:detecting_spoof_mechanisms}.

\autoref{fig:mechanism} shows how many VPs
  see non-anycast (injection or proxy, lightest area, on the bottom)
  or anycast spoofing (darkest, on top),
  from  2014-02 to 2020-05.

We see that non-anycast (injection or proxy)
  is by far the most popular spoofing mechanism, 
  accounting for 
  87\% to 100\% of the VPs that see spoofing.
We believe that non-anycast methods are popular because they do not involve
  routing, able to target at specific group of users;
  they can be deployed as a ``bump-in-the-wire''.
Anycast is the least popular one, since the anycast catchment relies on BGP, spoofers may not precisely control who to spoof.

We see 2 VPs
  that see alterations between overt spoofing and authentic replies,
  often with timeouts in between.
We speculate these VPs may have a mechanism that sometimes fails, e.g. a slow DNS injection, or site change between the authoritative or the third-party anycast site.
\PostSubmission{Should we look at these odd VPs some more?  Or not because they're jsut wierd?}

\subsection{Does Spoofing Speed Responses?}

Finally, we examine if spoofing provides faster responses
  than authoritative servers, since most of our identifiable spoofing parties are ISPs.

 \begin{figure}
 	\begin{center}
 		\includegraphics[width=0.85\columnwidth]{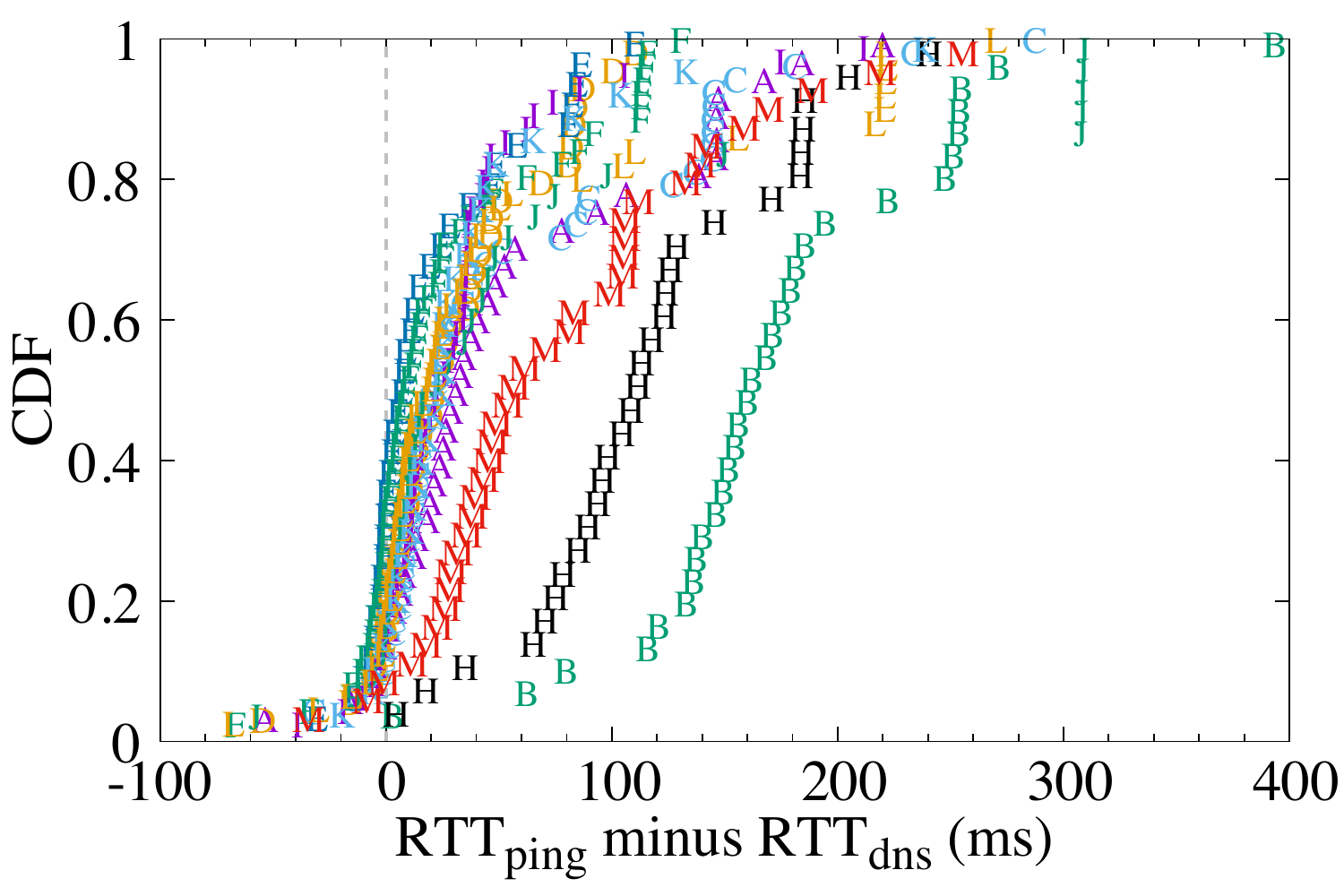}
 	\end{center}
 	\caption{CDF of $RTT_{ping}$ minus $RTT_{dns}$ from spoofed VPs on 2019-08-24}
 	\label{fig:enhance}
 \end{figure}

For each overt spoofer,
  we compare the median values of DNS response time with ping-time to the authoritative root on 2019-08-24.
In \autoref{fig:enhance}, we see that spoofing is almost always faster:
there can be about 15\% of all VPs that see equal or worse latency performance in spoofed answers.
This result is consistent with spoofing occurring near the VP.
In general, the amount of performance improvement is the inverse of the
  size of root letter's anycast footprint.
Letters with more anycast sites see less improvement,
  while for letters with only a few anycast sites (e.g. H-root, B-root), spoofing tends to be much faster.
This result is as one would expect for anycast latency~\cite{Schmidt17a},
  and is consistent with the statement that overt spoofers are
  improving user performance.
Except A-, B- , H-, and M-root, we also see that half of the VPs only see less than 20\,ms latency improvement from spoofers.
This shows even though spoofers improve performance but half of the VPs may still be good without them.

\section{Validation}
\label{sec:validation}

In this section, we validate the spoofing detection by the fact of whether the query has an answer from the authoritative B-root server or not.
First, we show our detection method can promise a true positive rate over 0.96.
Second, we show that other than the spoofing we detected, there are about 13 (0.14\%) VPs that may experience covert delayer over their DNS queries, and most of the cases, the DNS reply is slower than the Ping reply.
Third, we show that in recent days proxy is far more popular than injection is used, packets of about 98\% of spoofed queries are dropped on 2019-01-10.

\subsection{Validation Methodology}

Our spoof detection looks at traffic from VPs as DNS clients \autoref{sec:methodology}.
We validate it by looking at the destination side,
  from the authoritative server.
We expect queries from VPs that are intercepted and spoofed to not reach the server,
  while regular queries will appear in server traffic
  (unless there is packet loss or timeout).
For DNS injection, we expect the query to reach the server
  and two replies to return to the VP (first from the injector, then from the authoritative).

To validate our spoof detection we use
  server-side data from one week (2019-01-10 to -16, the only week available) of B-Root~\cite{broot_validate}.
That dataset uses host-only anonymization
  where the low-8 bits of the IPv4 address are scrambled,
  so we look for matches that have the same query type (DNS) and field name (hostname.bind)
  from the same IPv4 /24 prefix as the public address of each VP.
We also require that the timestamps are within 4 minutes as the RIPE Atlas querying interval.
(There are always multiple queries per second, so we cannot match by timestamp only.).
We use RIPE queries that are made directly to B-Root's IP address,
  not querying through the VP's recursive resolvers.

We apply our detection method during the same week 
  to match the B-Root dataset period.
Following \autoref{sec:dataset},
  we select four random full hour, each starting with a random offset (smaller than 3600 seconds) over the day
  from each day of the week.
  We evaluate each full hour.
We compare
queries sent from the RIPE Atlas VP
  that get a response or timeout
  to those seen at B-Root.
For each VP in the hourly window where some queries timeout and some succeed,
  we examine only successes.
With all queries timeout we classify that VP as timed-out.
We classify a timeout or spoof as correct if it does not show up at B-Root,
and any other query as correct if it does appear in B-Root traffic.

\acsacFix{B3:the paper does not provide experimental results on false positives. Response: our data cannot provide false positive because injection would leads to the results of ``false positive" but they are not actually false,. I added the following para.-Lan20201026}

A false positive is a query that is detected as spoofed
  where we can see that it actually reaches B-Root.
If a query does not reach B-Root and receives an answer,
  this query is definitely spoofed (a true positive).
Because of DNS injection, though,
  a spoofer may reply quickly to a query,
  but allow it to proceeds to B-Root
  where it then generates a second reply.
The scenario of DNS injection means that we cannot get a definitively
  count of false positives spoof detections,
  even with server-side data.
These potential false positives therefore place an upper bound on
  the actual false postive rate of spoofers;
    that upper bound is 0.02 ($1-0.98$) in \autoref{tab:b-root}.

\subsection{Validation of Overt Spoof Detection}

We first verify detections of overt spoofers (\autoref{sec:overt}).
We expect queries that see overt spoofing to \emph{not}  reach B-Root.
Since overt spoofing is obvious with atypical server IDs, we expect a high true positive rate.

\begin{table}
	\centering
	\begin{tabular}{l@{}rrr}
		&\multicolumn{3}{c}{\textbf{2019-01-10T03:52:49Z}}\\
		&sent&received& true positive rate\\
		active VPs & 8981 &8449& -\\
		\hspace{10pt}timeout & 241 & 47 & $\geq$0.81 \\
		\hspace{10pt}spoofed & 142 & 3 & $\geq$ 0.98 \\
		\hspace{20pt}non-anycast & 140 & 3 &  $\geq$0.98\\
		\hspace{20pt}anycast &2& 0 & 1\\
		\hspace{10pt}not spoofed & 8598 &8399& - \\
	\end{tabular}
	\caption{How many queries reach B-root based on spoof detection, for a sample hour.}
	\label{tab:b-root}
\end{table}
        

\begin{table}
	\centering
	\begin{tabular}{l|r r r r}
		& \multicolumn{4}{c}{\textbf{True-Positive Rate}}\\
		&\textbf{range} & \multicolumn{3}{c}{\textbf{quantile}}  \\
		\textbf{detection}&[min, max] & $0.25^{th}$ & $0.50^{th}$ & $0.75^{th}$\\
		\hline
		timeout & [0.79, 0.84]&0.8071&0.8198&0.8249\\
		spoof & [0.96, 0.99] & 0.9719 & 0.9787 & 0.9859\\
		not spoof &[0.90, 0.99]&0.9138&0.9297&0.9534\\
	\end{tabular}
\caption{The range of true positive rate of spoof detection from 2019-01-10 to 2019-01-16}
\label{tab:range}
\end{table}

\autoref{tab:b-root} shows a representative hour (other sample hours are similar),
  and \autoref{tab:range} shows the range of true positive rates for all 28 sample hours over the week.

The week of samples in \autoref{tab:range} shows that
  spoofing detection is accurate,
  with a true positive rate consistently around 0.97.
Examining a sample hour starting at 2019-01-10 3:52:49 GMT in \autoref{tab:b-root},
  142 of the 8981 VPs see spoofing.
For almost all VPs that see spoofing (139 of the 142),
  their queries do not arrive at B-Root, making a true positive rate over 0.98.
For mechanism, 140 of the 142 VPs experience either proxy or injection,
  and only 3 out of 140 VPs reached B-root, suggesting potentially DNS injection (the proxy drops the packets, so the query cannot reach B-root).
The rest 2 VPs suggest third-party anycast, and we confirm their queries
  are not seen at B-Root.

When examining VPs that timeout,
  the true positive fraction is around 0.82, with a wider range from 0.79 to 0.84 (see \autoref{tab:range}).
Some queries that timeout at the VP still reach B-root.
It is possible that a query reached B-Root and is answered,
  but the VP still timed out,
    perhaps because the reply was dropped by a third party.
The timeout default of RIPE Atlas probe is 5\,s.
In our example hour (\autoref{tab:b-root}), we see that out of 241 VPs that timeout, 47 of them has queries reach B-root, making a true positive rate of 0.81.

There are 199 VPs (about 2\%) VP in \autoref{tab:b-root} that neither were spoofed nor timeout,
  but for which we did not find a match on the B-Root side. 
It is possible the metadata of the IP address of those VPs is outdated or those VPs is multi-homed,
  so their queries arrive at B-Root from an IP address we do not know about.

\subsection{Validation of Covert Delayers}

\begin{table}
	\centering
	\begin{tabular}{l@{}rrr}
		&\multicolumn{3}{c}{\textbf{2019-01-10T03:52:49Z}}\\
		&detected&received& $RTT_{dns} - RTT_{ping}$\\
		covert-delayers & 13 & 13& -\\
		\hspace{10pt}$RTT_{dns} > RTT_{ping}$ & 12 & 12& 40.52ms\\
		\hspace{10pt}$RTT_{dns} \leq RTT_{ping}$ & 1& 1 & -10.25ms \\

	\end{tabular}
	\caption{Covert delayer validation: how many reached B-Root,
            with the mean difference for each DNS or ICMP faster.}
	\label{tab:covert_delayer}
\end{table}

We now examine covert delayers (\autoref{sec:covert}).
\autoref{tab:covert_delayer}
  shows analysis from one sample hour
  (other hours were similar).

First, we see that in all cases with a large delay,
  the queries \emph{does} get through to B-Root.
We originally expected differences in time indicated a covert spoofer,
  but these networks are passing the query to the authoritative server
  and not interfering with it.

However, we see there is a very large delay for the DNS replies.
Most of the time (12 of 13 cases) DNS is longer than ICMP,
  and the median difference is 40\,ms.
This consistent, large delay suggests that this difference is not just
  queueing delay or other noise in the network, and it is possible that a third-party is processing the traffic.

Although we do not have server-side data for other letters,
  we do see that about one-third of VPs that experience covert-delaying for B-Root also see covert-delaying
  with at least one other letter.

Finally, in one case we see a 10\,ms delay of ICMP relative to DNS.
it is possible that this delay is due to a router processing ICMP on the slow path.

\subsection{Non-Anycast Mechanism: Proxy or Injection?}
\label{sec:proxyorinjection}

Server-side data also allows us to distinguish DNS proxies
  from DNS injection.
DNS injection will respond quickly to the query while letting
  it pass through to the authoritative server (on-path processing) ,
  while a DNS proxy will intercept the query without passing it along
  (in-path processing).

In  \autoref{tab:b-root},
  shows that we see 139 out of 142 (98\%) of VPs that detected as spoofed never reach B-root, suggesting a DNS proxy instead of injection.
The remaining 3 VPs (only 2\%)
  are likely using DNS injection.
(Unfortunately we cannot confirm injection with a double reply at the receiver
  because we cannot modify the RIPE Atlas software.)

\section{Related Work}

Our work is inspired by prior work in
  improving DNS security, anycast location-mapping, and DNS spoofing detection.

Several groups have worked to improve or measure DNS security.
DNSSEC provides DNS integrity~\cite{Eastlake99b}.
Recent work of Chung et al.~\cite{chung2017longitudinal} shows under-use and mismanagement of the DNSSEC in about 30\% domains.
This work indicates that securing DNS involves actions of multiple parties.
Several groups explored DNS privacy and security,
  suggesting use of TLS to improve privacy~\cite{Zhu15b},
and methods to counter injection attacks~\cite{Duan12a}.
Others have identified 
  approaches to hijack or exploit DNS security~\cite{website2016,weaver2009detecting,vissers2017wolf,son2010hitchhiker},
  or studied censorship and multiple methods to spoof DNS~\cite{Gill15a,Ensafi:2015}.
Our work considers a narrower problem, and explores it over more than six years of data:
  we study who, where and how DNS spoofing occurs.
Our work complements this prior work by motivating deployment of defences.
\comment{I cite Liu's work here-Lan20201119}
\comment{good, you summarize them and say where we agree.  but what do we do that they dind't already do?  i.e., why publish our paper sinc ethey already published? ---johnh 2020-11-23}
\comment{I added another sentence saying the differences-Lan20201123 }
\comment{thanks.  I rephrased it to be shorter. ---johnh 2020-11-23}
Liu et al. looks at DNS spoofing when users use public DNS servers~\cite{liu2018answering}.
This work points out that interception happens about 10 times more than injection in TLD DNS queries.
This finding agrees with our conclusion that proxies (interception) account for nearly all spoofing we see.
Our work goes beyond their work to characterize who third-parties are,
  and to study longitudinal data.

Several groups have studied the use of anycast and how to optimize performance.
Fan et al.~\cite{fan2013} used traceroute and open DNS resolvers
  to enumerate anycast sites, as well as Server ID information.
They mention spoofing, but do not study it in detail.
Our work also uses Server ID
   and traceroute to study locations,
  but we focus on identifying spoofers, and how and where they are.
Other prior work uses Server ID to
  identify DNS location
  to study DNS or DDoS~\cite{Moura16b,Schmidt17a,wei2017does}.

Work of Jones et al.~\cite{jones2016detecting} aims to find
  DNS proxies and unauthorized root servers. 
They study B-root because it was unicast at the time,
  making it easy to identify spoofing.
Our work goes beyond this work to study spoofing over all 13 letters
  over more than six years,
  and to identify spoofing mechanisms.

Closest to our work,
  the Iris system is designed to detect DNS manipulation globally~\cite{paul2017censorship}.
They take on a much broader problem, studying all methods of manipulation across all of the DNS hierarchy,
  using open DNS resolvers.
Our work considers the narrower problem of spoofing
  (although we consider three mechanisms for spoofing),
  and we study the problem with active probing from RIPE Atlas.
Although our approach generalizes, we analyse only the DNS Root.

Finally,
  we recently became aware that Wessels is studying currently spoofing at the DNS Root
  with Ripe Atlas~\cite{Wessels19a}.
His work is not yet generally available and is in progress,
  but to our knowledge, he does not look at spoofing mechanisms
  and has not considered six years of data.

\section{Conclusion}

This paper developed new methods to detect overt DNS spoofing and some covert delayers,
  and to identify and classify parties carrying out overt spoofing.
In our evaluation of about six years of spoofing at the DNS Root,
  we showed that spoofing is quite rare,
  affecting only about 1.7\% of VPs.
However, spoofing is increasing, growing by more than $2 \times$ over more than six years.
We also show that  spoofing is global, although more common in some countries.
By validating using logs of authoritative server B-root, we prove that our detection method has true positive rate of at least 0.96.
Finally, we show that proxies are a more common method of spoofing
  today than DNS injection.

We draw two recommendations from our work.
First, 
  based on the growth of spoofing, we recommend that operators regularly
  look for DNS spoofing.
Second, interested end-users may wish to watch for spoofing using
  our approach.

{
	\bibliographystyle{abbrv}
	\bibliography{paper}
}

\ifistechreport
\appendix

\section{Does Spoofing Speed Responses?}
	\label{sec:speed_appendix}

\comment{John, please have a look--Lan20190510}

\comment{this section is better in that you finally start with a question.
But you need to tie it into your other sections. 
Why are you asking THIS question NOW, and how does it relate to your OTHER questions? ---johnh 2019-05-07}
\comment{done-Lan20190510}

We are now curious how much the spoofing helps the performance of DNS queries, since we now know that most identifiable spoofers are ISPs or network providers try to provider better services using spoof in \autoref{sec: who}.

We compare the RTTs in the spoofed DNS responses and RTT of pinging each root letter to see if the spoofed responses are always faster than the ping.
For each spoofed VP, we compute the mean of its ping RTTs and its spoofed DNS RTTs in the same two-hour window (Both reponses are at the same frequency). 

Spoofing helps most VPs shorten the latency. 
As we can see in \autoref{fig:spoof_help},
majority of the CDF dots has a positive values,
showing the spoofed response is much faster than pings to the actual root servers.
For some letters as H, M, and B-root, the latency boost can be as high as 200 to 300 ms.

Spoofing sometimes won't help the latency that much, or is even slower than the authentic reponses.
Except B-root and H-root, other letters all have CDFdots that are negative, although not so many.
This means for a few VPs, it is even faster for them to use the authentic DNS instead of the spoofing provided by the ISPs.
When we look at where spoofing RTT  is 10 ms faster than Ping (x = 10ms), we see except A, B, H ,M root, other letters have 20\% or more VPs only get 10ms or less performance gain from spoofing.
We think it is reasonable that some users are better off with authentic servers.
Because authentic root DNS are built on anycast, which tries to provide better location-coverage for shorter latency. 
It is also fair that for
B- and H- root, spoofing is always faster than authentic, 
since they are the only two sites that have only two location covered globally.
They cannot answer worldwide VPs in a quicker-than-spoof way limited by physical distance. 

Spoofers may have a heavy underperformance due to overload or other reasons in rare cases.
We omit two VPs in \autoref{fig:spoof_help}.
For one probe located in Russia, and on C-root and D-root, spoofer replies DNS query in more than 2000 ms while ICMP ping from authentic DNS only spent about 40 ms for both roots. 
We think this suggest single-node spoofers may be as robust as anycast authentic root services.

\begin{figure}
	\begin{center}
		\includegraphics[width=0.90\columnwidth]{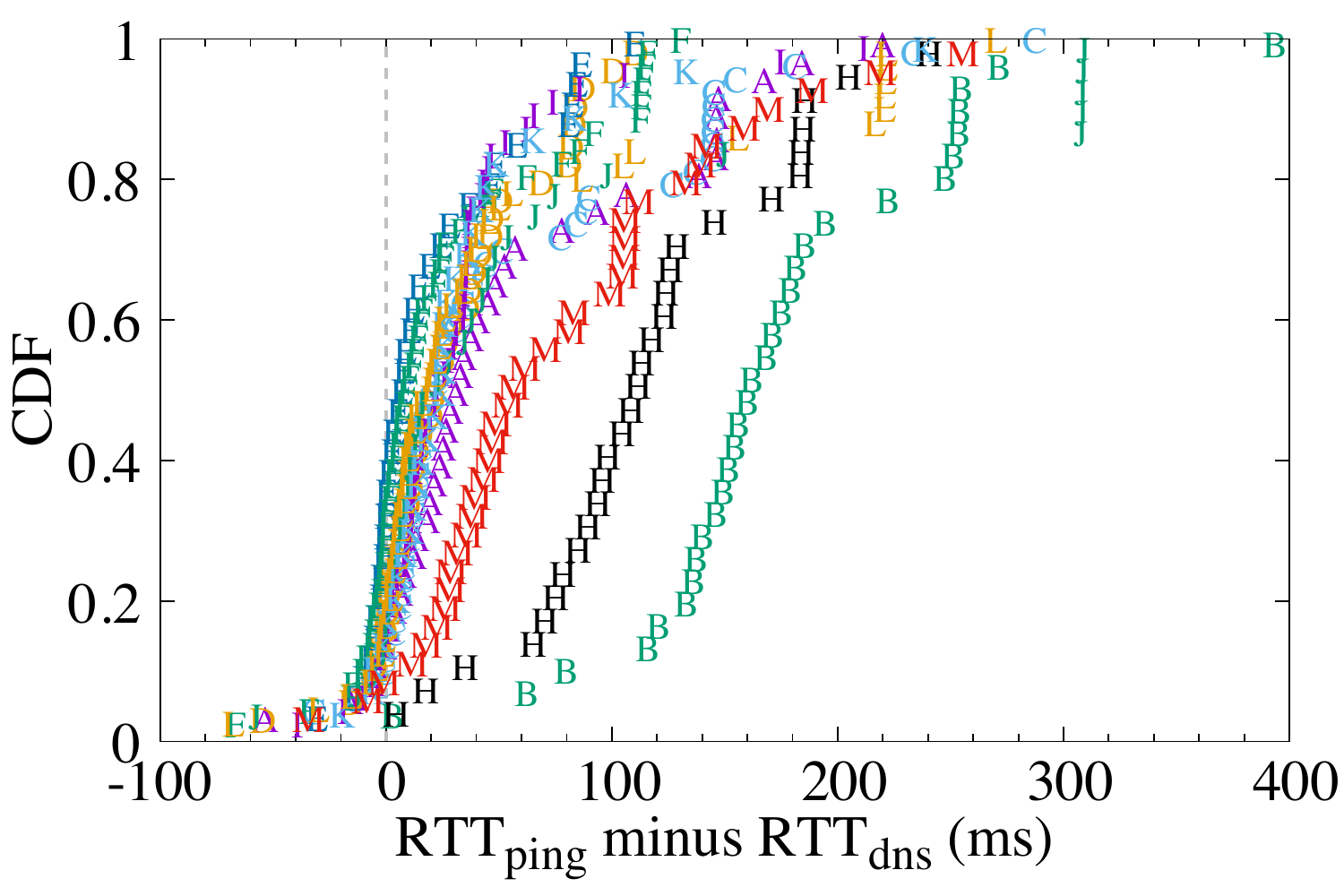}
	\end{center}
	\caption{CDF of the latency gain from spoofed answer in 2019-01. This graph does not include G-root (not answering ICMP Ping). }
	\label{fig:spoof_help}	
\end{figure}

\fi

\end{document}